\documentclass[
final,
reprint,
showkeys,
aps,
pre,
floatfix,
]{revtex4-2}

\usepackage{nameref}
\usepackage[dvips]{graphicx}
\usepackage[utf8]{inputenc}
\usepackage{multirow,booktabs}
\usepackage{amsmath,amssymb,latexsym,MnSymbol,bm}
\usepackage{blkarray}

\usepackage[table]{xcolor}
\usepackage{xcolor}
\usepackage[%
  colorlinks=true,
  urlcolor=blue,
  linkcolor=blue,
  citecolor=blue
]{hyperref}
\usepackage{blindtext}
\usepackage{scalerel,stackengine}
\usepackage{enumitem}

\begin{document}

\title{Mapping images into ordinal networks}

\author{Arthur\ A.\ B.\ Pessa}
\email{arthur\_pessa@hotmail.com}
\affiliation{Departamento de F\'isica, Universidade Estadual de Maring\'a -- Maring\'a, PR 87020-900, Brazil}

\author{Haroldo\ V.\ Ribeiro}
\email{hvr@dfi.uem.br}
\affiliation{Departamento de F\'isica, Universidade Estadual de Maring\'a -- Maring\'a, PR 87020-900, Brazil}

\date{\today}

\begin{abstract}
An increasing abstraction has marked some recent investigations in network science. Examples include the development of algorithms that map time series data into networks whose vertices and edges can have different interpretations, beyond the classical idea of parts and interactions of a complex system. These approaches have proven useful for dealing with the growing complexity and volume of diverse data sets. However, the use of such algorithms is mostly limited to one-dimension data, and there has been little effort towards extending these methods to higher-dimensional data such as images. Here we propose a generalization for the ordinal network algorithm for mapping images into networks. We investigate the emergence of connectivity constraints inherited from the symbolization process used for defining the network nodes and links, which in turn allows us to derive the exact structure of ordinal networks obtained from random images. We illustrate the use of this new algorithm in a series of applications involving randomization of periodic ornaments, images generated by two-dimensional fractional Brownian motion and the Ising model, and a data set of natural textures. These examples show that measures obtained from ordinal networks (such as average shortest path and global node entropy) extract important image properties related to roughness and symmetry, are robust against noise, and can achieve higher accuracy than traditional texture descriptors extracted from gray-level co-occurrence matrices in simple image classification tasks.
\end{abstract}

\maketitle

\section{Introduction}\label{sec:intro}

In the last two decades, network science has established itself as a vibrant and successful field of research~\cite{vespignani2018twenty}. The fact that diverse complex systems are accurately described as sets of vertices and edges~\cite{dorogovtsev2010lectures,albert2002statistical} combined with advancements in data acquisition and processing have led to the widespread application of networks to an 
immense variety of systems from biological, sociological and technological origins. In addition to these more established works, recent developments in network science have originated new and more abstract forms to define complex networks. Vertices and edges of these more abstract networks usually do not represent constituents and interactions of a system; instead, algorithms mapping objects into networks ascribe different meanings to these basic network components~\cite{zanin2016combining,zou2019complex}.

Algorithms designed to map time series into networks are a particularly important class of such networks~\cite{zou2019complex} which have been used to characterize time series of chaotic and stochastic nature obtained from simulations and experimental data. These approaches are collectively known as time series networks and the most prominent examples include visibility graphs~\cite{lacasa2008from}, recurrence networks~\cite{marwan2009Complex} and ordinal networks~\cite{small2019complex}. Visibility graphs and its variants~\cite{lacasa2008from,luque2009horizontal,lacasa2012time, bezsudnov2014from,bianchi2017multiplex} map each time series observation into a network vertex, and any two vertices are connected if their respective values in the time series satisfy a visibility condition~\cite{lacasa2008from}. Recurrence networks, on its turn, are a reinterpretation of recurrence plots~\cite{eckmann1987recurrence}, a graphic tool developed in the context of nonlinear time series analysis~\cite{kantz2004nonlinear,bradley2015nonlinear}. Vertices of recurrence networks represent a set of states obtained from small segments of time series, and edges connect vertices (pairs of states) that are similar according to a distance metric~\cite{marwan2009Complex}. Ordinal networks were proposed more recently but also originate from the study of nonlinear time series, specifically from a successful symbolization approach due to \citeauthor{bandt2002permutation}~\cite{bandt2002permutation}. Nodes of ordinal networks represent ordering patterns (or permutation symbols) associated with time series slices, and links are drawn based on the succession of these symbols in the time series~\cite{small2019complex, mccullough2015time,pessa2019characterizing,olivares2020contrasting}.

While the use of algorithms mapping time series into networks is increasingly gaining popularity among the scientific community (see \citeauthor{zou2019complex}~\cite{zou2019complex} for a recent review), few efforts have been dedicated to extending these ideas to higher-dimensional data such as images. Most works on this direction have focused on representations of images as spatial networks where pixels (or groups of pixels) are mapped into nodes with links reflecting distance and color similarity among them. These methods have proved useful for edge detection~\cite{wu2015image}, texture extraction~\cite{xu2015complex}, image segmentation~\cite{cuadros2012segmentation}, among other practical applications~\cite{backes2013texture,gonccalves2015complex,backes2009complex}. However, there have been even fewer attempts beyond these more regular network representations. Notable exceptions include the works of \citeauthor{xiao2014row}~\cite{xiao2014row} on the degree distributions of row-column visibility graphs obtained from fractal landscapes, and \citeauthor{lacasa2017visibility}~\cite{lacasa2017visibility,iacovacci2019visibility} that analyzed visibility networks mapped from bidimensional random and chaotic data and tackled problems of image processing and classification. As processes for extracting meaningful information from images are ubiquitous across science, extending and proposing approaches for mapping images into complex networks have great potential to contribute with new image quantifiers derived from well-established network metrics.

Here we present an extension of the ordinal network framework allowing the representation of images as complex networks. We describe intrinsic connectivity constraints of ordinal networks inherited from the symbolization process and determine the exact form of ordinal networks mapped from completely noisy images. By mapping images obtained from periodic ornaments, two-dimensional fractional Brownian motion, and the Ising model into ordinal networks, we illustrate the use of this new approach to identify and describe these systems with network-related metrics. We apply our method to characterize a data set of real-world images, where we show that ordinal networks are capable of distinguishing different types of textures and identifying image symmetries. We further demonstrate that ordinal network measures are robust against noise addition and display higher accuracy than traditional texture descriptors in simple image classification tasks.

The rest of this paper is organized as follows. In Section~\ref{sec:methods}, we briefly revisit the ordinal network framework before generalizing it to two-dimensional data. Next, we investigate connectivity properties of ordinal networks (Section~\ref{sec:constraints}), the exact form of ordinal networks obtained from random data (Section~\ref{sec:random}), and noisy-periodic ornaments (Section~\ref{sec:ornaments}). Applications involving fractal and Ising surfaces are presented in Sections~\ref{sec:fbm} and \ref{sec:ising}, real-world images (Brodatz textures) are investigated in Section~\ref{sec:brodatz}, and simple image classification tasks comparing the performance of ordinal network with traditional texture descriptors are presented in Section~\ref{sec:comparing}. Finally, Section~\ref{sec:conclusion} concludes our work.

\section{Methods}\label{sec:methods}

The representation of time series as ordinal networks has direct relations to permutation entropy, a successful time series complexity measure~\cite{bandt2002permutation}. Specifically, ordinal networks use the same approach introduced by \citeauthor{bandt2002permutation}~\cite{bandt2002permutation} (see also \citeauthor{bandt2007order}~\cite{bandt2007order}) to partition a time series in small segments and associate a permutation symbol (or an ordinal pattern) with each part according to the relative amplitude of the time series values~\cite{pessa2019characterizing}. Our extension of the ordinal network algorithm for two-dimensional data is inspired by a generalization of permutation entropy to image data proposed in Ref.~\cite{ribeiro2012complexity} that proved useful for investigating liquid crystals~\cite{zunino2016discriminating,sigaki2019estimating} and art paintings~\cite{sigaki2018history}.

By following Ref.~\cite{ribeiro2012complexity}, we start by considering a two-dimensional array $\{y_i^j\}_{i=1,\dots,N_x}^{j=1,\dots,N_y}$ of size $N_x {\times} N_y$, where the elements $y_i^j$ may represent pixels of an image. Next, we divide this array into sliding partitions of size $d_x$ by $d_y$ defined as
\begin{equation}\label{eq:matrix_partition}
    w_{s}^{t} = 
\begin{pmatrix}
    y_{s}^{t} & y_{s}^{t+1}   & \dots & y_{s}^{t+(d_y-1)} \\[.5em]
    y_{s+1}^{t} & y_{s+1}^{t+1}   & \dots & y_{s+1}^{t+(d_y-1)} \\
    \vdots & \vdots & \ddots & \vdots\\
    y_{s+(d_x-1)}^{t} & y_{s+(d_x-1)}^{t+1}   & \dots & y_{s+(d_x-1)}^{t+(d_y-1)} \\
\end{pmatrix},
\end{equation} 
where the indices $s = 1,\dots, n_x$ and $t = 1,\dots, n_y$, with $n_x = N_x-d_x+1$ and $n_y = N_y-d_y+1$, cover all $n_x n_y$ possible sliding partitions. The values of $d_x$ and $d_y$ are the two parameters of the approach and represent the horizontal and vertical embedding dimensions~\cite{ribeiro2012complexity}. We then flatten these two-dimensional partitions line by line as
\begin{equation}
\begin{split}
    w_{s}^{t} =  & \left( y_{s}^{t}, y_{s}^{t+1}, \dots, y_{s}^{t+(d_y-1)}, \right.\\
                 & ~~y_{s+1}^{t}, y_{s+1}^{t+1},\dots,y_{s+1}^{t+(d_y-1)},\dots,\\
                 & \left. ~y_{s+(d_x-1)}^{t}, y_{s+(d_x-1)}^{t+1}, \dots,  y_{s+(d_x-1)}^{t+(d_y-1)}\right)\,,
\end{split}
\end{equation}
to investigate the ordering of its elements. Because this procedure does not depend on the partition location (that is, $s$ and $t$), we can simplify the notation and rewrite the flattened partition as 
\begin{equation}\label{eq:partition}
    w = \left(\tilde{y}_0, \tilde{y}_1, \dots, \tilde{y}_{d_x d_y-2}, \tilde{y}_{d_x d_y-1}\right)\,,
\end{equation}
where $\tilde{y}_0 = y_{s}^{t},~\tilde{y}_1 = y_{s}^{t+1}$, and so on. 

Under this notation, the symbolization procedure consists in evaluating the permutation $\Pi = (r_0, r_1, \dots, r_{d_x d_y-2}, r_{d_x d_y-1})$ of the index numbers $(0,1,\dots,d_x d_y{-}2,d_x d_y{-}1)$ that sorts the elements of the flattened partition in ascending order, that is, the index numbers resulting in $\tilde{y}_{r_{0}} \leq \tilde{y}_{r_{1}} \leq \dots \leq \tilde{y}_{r_{d_x d_y-2}} \leq \tilde{y}_{r_{d_x d_y-1}}$. In case of draws, we maintain the occurrence order of the elements in $w$, that is, $r_{j-1} < r_{j}$ if $\tilde{y}_{r_{j-1}} = \tilde{y}_{r_{j}}$ for $j=1,\dots,d_x d_y-1$~\cite{cao2014detecting}. To illustrate this procedure, suppose we have $d_x=d_y=2$ and the partition matrix 
$w=
\begin{pmatrix}
    4 & 5 \\
    2 & 2
\end{pmatrix}.$ The corresponding flattened array is $w=(4,5,2,2)$, and so $\tilde{y}_0=4,~\tilde{y}_1=5,~\tilde{y}_2 = 2$ and $\tilde{y}_3=2$. Because $\tilde{y}_{2} \leq \tilde{y}_{3} \leq \tilde{y}_0 \leq \tilde{y}_1$, the permutation $\Pi = (2,3,0,1)$ is the one that sorts the elements of the partition $w$. 

After carrying out the symbolization procedure over the entire data array, we construct another array $\{\pi_s^t\}_{s=1,\dots,n_x}^{t=1,\dots,n_y}$ containing the permutation symbols associated with each sliding partition $w_s^t$. By using this new array, we can calculate the relative frequency $\rho_i(\Pi_i)$ of each possible permutation $\Pi_i$ defined as
\begin{equation}
    \rho_i(\Pi_i) = \frac{\text{number of partitions of type $\Pi_i$ in } \{\pi_s^t\}}{n_x n_y}\,,
\end{equation}
where $i=1,\dots,(d_x d_y)!$ and $(d_x d_y)!$ is the total number of possible permutations that can occur in the original data array. Having these relative frequencies, we construct the probability distribution $P=\{\rho_i(\Pi_i)\}_{i=1,\dots,(d_x d_y)!}$ of ordinal patterns and estimate the two-dimensional version of the permutation entropy~\cite{ribeiro2012complexity}
\begin{equation}
    H = -\sum_{i=1}^{(d_x d_y)!} \rho_i(\Pi_i)\log \rho_i(\Pi_i)\,,
\end{equation}
where $\log(\dots)$ stands for the base-$2$ logarithm. This generalized version of the permutation entropy recovers the one-dimensional case (time series data or $N_y=1$) by setting $d_y = 1$ and properly choosing $d_x$. It is worth noticing that the embedding dimensions $d_x$ and $d_y$ must satisfy the condition $(d_x d_y)! \ll N_x N_y$ in order to obtain a reliable estimate of the probability distribution $P=\{\rho_i(\Pi_i)\}_{i=1,\dots,(d_x d_y)!}$~\cite{bandt2002permutation,ribeiro2012complexity}. 

\begin{figure*}[!ht]
\centering
\includegraphics[width=1\linewidth]{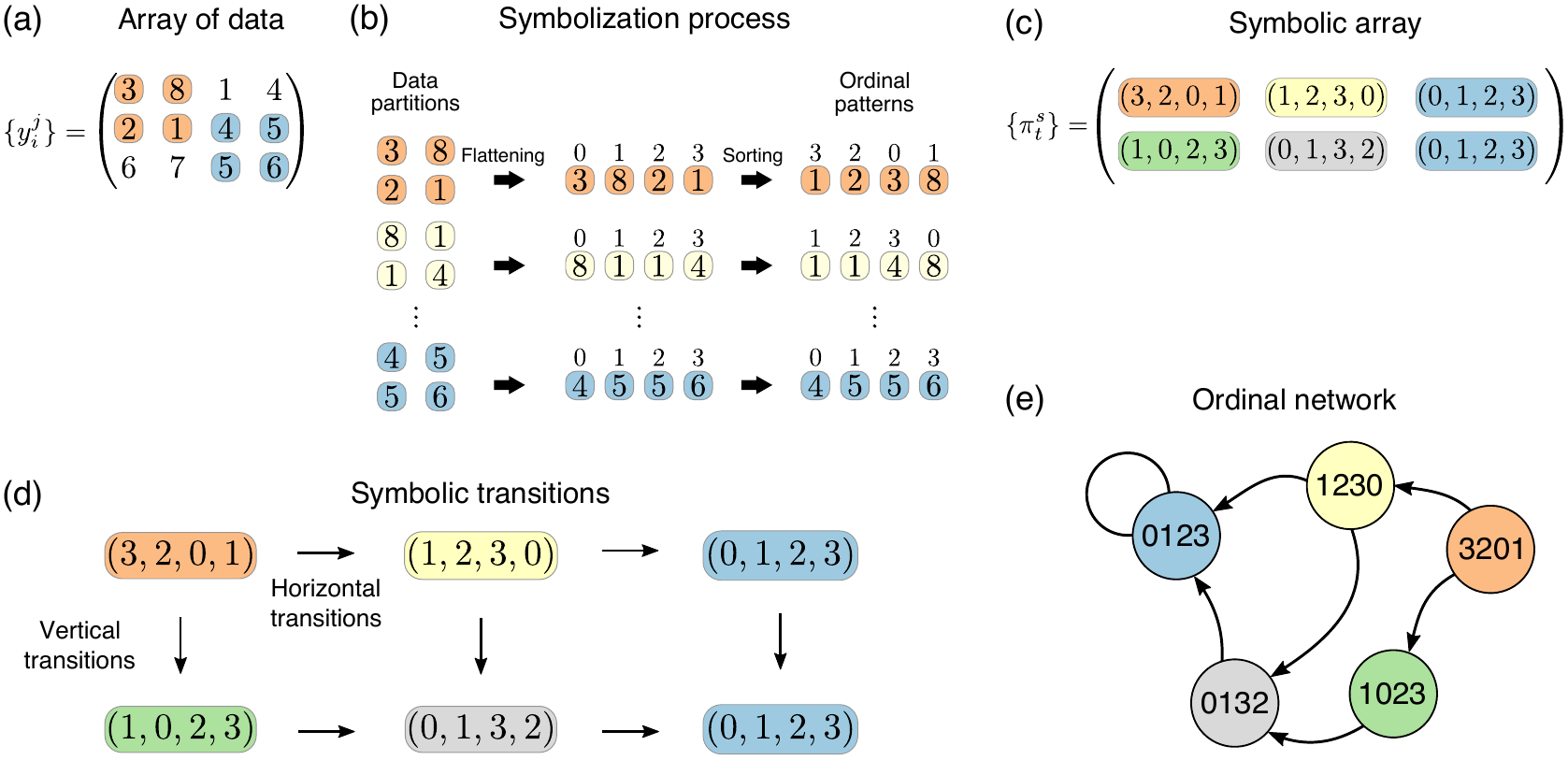}
\caption{Mapping two-dimensional data into ordinal networks. (a) A small illustrative array of data $\{y_i^j\}_{i=1,\dots,N_x}^{j=1,\dots,N_y}$ of size $N_x=3$ and $N_y=4$. (b) Illustration of the symbolization process applied to the data with embedding dimensions $d_x=d_y=2$. This process essentially consists in evaluating the ordering of the values within each data partition. (c) Array with the symbolic sequences (or permutation patterns) $\{\pi_s^t\}_{s=1,\dots,n_x}^{t=1,\dots,n_y}$ ($n_x=3$ and $n_y=2$) obtained from the original data. (d) All first-neighbor transitions (or vertical and horizontal successions) among ordinal patterns occurring in the symbolic array. (e) Representation of data array as an ordinal network. In this example, all permutation successions occur only once and so the network edges have all the same weight. Self-loops [as in the permutation $\Pi=(0,1,2,3)$] emerge whenever a permutation pattern is adjacent to itself in the symbolic array.}
\label{fig:method}
\end{figure*}

To generalize the concept of ordinal networks to two-dimensional data, we use the symbolic array $\{\pi_s^t\}_{s=1,\dots,n_x}^{t=1,\dots,n_y}$ obtained from the previous discussion. As in the one-dimensional case, we consider each unique permutation symbol $\Pi_i$ [$i=1,\dots,(d_x d_y)!$] occurring in $\{\pi_s^t\}$ as a node of the corresponding ordinal network. Next, we draw directed edges between these nodes according to the first-neighbor transitions occurring in $\{\pi_s^t\}_{s=1,\dots,n_x}^{t=1,\dots,n_y}$, that is, we directly-connect the permutation symbols involved in all horizontal and vertical successions among ordinal patterns in the symbolic array ($\pi_{s}^{t} \to \pi_{s}^{t+1}$ and $\pi_{s}^{t}\to \pi_{s+1}^{j}$, with $s = 1,\dots,n_x-1$ and $t = 1,\dots,n_y-1$). The directed link between a pair of permutation symbols $\Pi_i$ and $\Pi_j$ is weighted by the total number of occurrences of this particular transition in the symbolic array. Thus, we can write the elements of the weighted-adjacency matrix representing the ordinal network as
\begin{equation}\label{eq:edge_weights}
    p_{i,j} = \frac{\text{total of transitions $\Pi_i \to \Pi_j$ in $\{\pi_s^t\}_{s=1,\dots,n_x}^{t=1,\dots,n_y}$}}{2n_x n_y-n_x-n_y}\,,
\end{equation}
where $i, j = 1,\dots,(d_x d_y)!$ and the denominator represents the total number of horizontal and vertical permutation successions in $\{\pi_s^t\}_{s=1,\dots,n_x}^{t=1,\dots,n_y}$. Figure~\ref{fig:method} illustrates the procedure for creating an ordinal network from a simple two-dimensional array of size $N_x=3$ and $N_y=4$.

In addition to more usual network metrics, the probabilistic aspects of nodes and edges in ordinal networks can also be quantified by entropy measures at the node level or for the whole network. Given an ordinal network vertex $i$ (associated with a permutation $\Pi_i$), the local node entropy~\cite{mccullough2007multiscale,pessa2019characterizing} is defined for this vertex as
\begin{equation}
    h_i = -\sum_{j\in\mathcal{O}_i}p'_{i, j}\log p'_{i,j},
\end{equation}
where $p'_{i,j} = p_{i,j}/\sum_{k\in\mathcal{O}_i}p_{i,k}$ represents the renormalized transition probability of transitioning from node $i$ to node $j$ (associated with the permutation $\Pi_j$), and $\mathcal{O}_i$ is the outgoing neighborhood of node $i$ (set of all edges leaving node $i$). The local node entropy $h_i$ quantifies the degree of determinism related to  permutation transitions at the node level. We have $h_i=0$ (deterministic case) when only one edge leaves node $i$, whereas $h_i$ is maximum if all edges leaving $i$ have the same weight (equiprobable case). At the network level, we can define the global node entropy as
\begin{equation}\label{eq:global_node_entropy}
    H_{\rm GN} = \sum_{i=1}^{(d_x d_y)!}p'_ih_i\,,
\end{equation}
where $p'_i = \sum_{j\in\mathcal{I}_i}p_{j,i}$ corresponds to the probability of transitioning to node $i$ from its incoming neighborhood $\mathcal{I}_i$ (in-strength of node $i$). If the original data array is large enough [$(d_xd_y)!\gg N_x N_y$], $p'_i$ converges to the probability of occurrence of permutation $\pi_i$, and $H_{\rm GN}$ corresponds to a weighted average of local node determinism throughout the network.

\section{Results}\label{sec:results}

\subsection{Connectivity constraints}\label{sec:constraints}

Having described our extension of the ordinal network algorithm for two-dimensional data, we start by investigating how properties of the symbolic array of permutations affect the connectivity of the resulting network. Similarly to what has been found for time series~\cite{pessa2019characterizing}, ordinal networks mapped from two-dimensional arrays also present restrictions that prohibit the existence of several edges among permutation symbols. These constraints result from the symbolization approach and the procedure used to partition the array of data; they emerge even when data is completely random.

To better illustrate this finding, we consider a data array $\{y_i^j\}_{i=1,\dots,N_x}^{j=1,\dots,N_y}$ and embedding dimensions $d_x=d_y=2$. Let us now suppose that the first partition matrix 
$w_1^1 = 
\begin{pmatrix} 
y_1^1 & y_1^2 \\ 
y_2^1 & y_2^2 
\end{pmatrix}$
is characterized by the permutation $\pi_1^1 = (2,1,3,0)$ and that the next horizontally-adjacent partition is
$w_1^2 = 
\begin{pmatrix} y_1^2 & y_1^3 \\ 
y_2^2 & y_2^3
\end{pmatrix}.$
We note that both partitions share the array elements $y_1^2$ and $y_2^2$. Thus, because $w_1^1$ is described by $\pi_1^1 = (2,1,3,0)$ (that is, $y_2^1 < y_1^2 < y_2^2 < y_1^1$), the permutation $\pi_1^2$ associated with $w_1^2$ must respect the inequality $y_1^2 < y_2^2$ imposed by $\pi_1^1$. Out of the $(d_x, d_y)!=24$ possible permutations, the previous constraint restricts $\pi_1^2$ to one among $12$ permutations in which the index number $0$ (corresponding to the position of $y_1^2$ in $w_1^2$) precedes the number $2$ (corresponding to the position of $y_2^2$). Vertically-adjacent partitions also present similar constraints. For instance, 
$
w_2^1 = 
\begin{pmatrix}
y_2^1 & y_2^2 \\
y_3^1 & y_3^2
\end{pmatrix}$
shares the array elements $y_2^1$ and $y_2^2$ with $w_1^1$, and the index number $0$ (position of $y_2^1$ in $w_2^1$) has to precede the ordinal number $1$ (index number of $y_2^2$ in $w_2^1$) in $\pi_2^1$ because of the inequality $y_2^1 < y_2^2$ expressed by $\pi_1^1$. Thus, $\pi_2^1$ is also constrained by $\pi_1^1$ to be one among $12$ ordinal patterns. These restrictions hold independently of the particular permutation corresponding to $\pi_1^1$, and for each permutation, there are only $12$ others that can immediately follow it (horizontally or vertically) when $d_x = d_y = 2$.

The same reasoning applies to permutation successions involving arbitrary values of the embedding dimensions $d_x$ and $d_y$. However, it is important to notice that the number of constraints in successions among permutations increases as adjacent partitions share a larger number of array elements. For instance, horizontally-adjacent partitions share two array elements while vertically-adjacent partitions share three array elements when $d_x=3$ and $d_y=2$. For these embedding dimensions, out of all $(3 \times 2)!=720$ possible permutations, once the permutation $\pi_s^t$ is set, there are only $30$ allowed ordinal patterns for its horizontal-neighbor permutation $\pi_s^{t+1}$ and $120$ allowed ordinal patterns for its vertically-adjacent permutation $\pi_{s+1}^t$. 

Given $d_x$ and $d_y$, the number of horizontal neighbors of a permutation is $C(d_x d_y, d_y)\times d_y!$, where $C(a,b)=\frac{a!}{b!(a-b)!}$. The binomial term represents the total of combinations between the number of elements inside a data partition ($d_x d_y$) and the number of horizontally shared elements between horizontally-adjacent partitions ($d_y$). The multiplication by $d_y!$ indicates that for each arrangement of the shared elements between two adjacent partitions, the non-shared elements can have $d_y!$ amplitude relations between themselves. Similarly, it follows that the number of allowed vertical neighbors is $C(d_x d_y, d_x)\times d_x!$. 

Once we know the set of allowed horizontal and vertical neighbors for a given permutation, the maximum number of outgoing edges in the vertex associated with this particular permutation is the intersection of these two sets. Interestingly, the sets of horizontal and vertical neighbors are not disjoint. Thus, the maximum number of outgoing edges depends on the ordinal pattern itself and is not simply the sum of the horizontal and vertical neighbors. For instance, the ordinal pattern $\Pi=(0,1,2,3)$ can have at most $16$ outgoing edges while $\Pi=(0,1,3,2)$ can have up to $20$ outgoing connections in an ordinal network.

By investigating the maximum number of allowed edges for every permutation node, we can find the maximum number of edges for the whole ordinal network. For embedding dimensions $d_x = d_y = 2$, we find that the ordinal network resulting from an arbitrary data array can have up to 24 vertices linked by 416 edges. As in the one-dimensional case~\cite{pessa2019characterizing}, the number of nodes and edges in ordinal networks increases dramatically with the embedding dimensions. For instance, an ordinal network can be formed by $720$ nodes and $104,184$ edges when $d_x=3$ and $d_y=2$. This imposes important limitations in the investigation of small data samples since the condition $(d_x d_y)! \ll N_x N_y$ is required to have a reliable estimate for the transition probabilities (edge weights). Most empirical investigations are therefore limited to values of embedding dimensions equal to two or three. However, recent investigations with time series indicate that the analogous condition imposed on permutation entropy can be considerably relaxed and still yield stable results and high accuracy in classification settings~\cite{cuesta2019embedded}. This result suggests that missing ordinal patterns~\cite{amigo2007true} carry important information about the system under investigation. We believe ordinal networks are likely to present a similar behavior, where missing ordinal patterns (network nodes) and missing transitions (network edges) could also be used for classification purposes. Thus, while it is common practice to impose the condition $(d_x d_y)! \ll N_x N_y$, the results of Ref.~\cite{cuesta2019embedded} indicate that one may consider the embedding dimensions $d_x$ and $d_y$ as tuning parameters in classification tasks.

\subsection{Random ordinal networks}\label{sec:random}
The constraints discussed in the previous section already indicate that ordinal networks emerging from completely random arrays are not random graphs. As it has been recently uncovered for time series~\cite{pessa2019characterizing}, ordinal networks resulting from random data (called random ordinal networks) have a non-trivial network structure. We now investigate how these findings generalize for two-dimensional random data. To do so, we consider an array of data $\{y_i^j\}_{i=1,\dots,N_x}^{j=1,\dots,N_y}$ sampled from a continuous probability distribution and embedding dimensions $d_x = d_y = 2$. We then extract the first two horizontally-adjacent partitions
$w_1^1 = 
\begin{pmatrix} 
y_1^1 & y_1^2 \\ 
y_2^1 & y_2^2 
\end{pmatrix}$
and 
$w_1^2 = 
\begin{pmatrix}
y_1^2 & y_1^3 \\
y_2^2 & y_2^3 
\end{pmatrix},$
and suppose that $y_2^1 < y_1^2 < y_2^2 < y_1^1$, that is, $\pi_1^1 = (2,1,3,0)$. To determine the transition probabilities (edge weights as defined in Eq.~\ref{eq:edge_weights}) from $\pi_1^1$ to other permutations, we need to find all possible permutations which can be associated to $\pi_1^2$ by evaluating the inequalities involving the values in $w_1^1$ and $w_1^2$. 

We can start by analyzing all possible amplitude relations between $y_1^3$ and the already established condition $y_2^1 < y_1^2 < y_2^2 < y_1^1$ to find:
\begin{equation}
\begin{split}
    i) \ y_2^1 < y_1^2 < y_2^2 < y_1^1 < y_1^3; \\
    ii) \ y_2^1 < y_1^2 < y_2^2 < y_1^3 < y_1^1; \\
    iii) \ y_2^1 < y_1^2 < y_1^3 < y_2^2 < y_1^1; \\
    iv) \ y_2^1 < y_1^3 < y_1^2 < y_2^2 < y_1^1 ;\\ 
    v) \ y_1^3 < y_2^1 < y_1^2 < y_2^2 < y_1^1.    
\end{split}
\end{equation}
Next, we include $y_2^3$ and analyze each possible amplitude relation for all the previous five conditions. This procedure leads to a total of 30 possible amplitude relations between the six elements contained in partitions $w_1^1$ and $w_1^2$: 
\begin{equation}\label{eq:all30inequalities}
\begin{split}
    i) \ y_2^1 < y_1^2 < y_2^2 &< y_1^1 < y_1^3 < y_2^3; \\
    ii) \ y_2^1 < y_1^2 < y_2^2 &< y_1^1 < y_2^3 < y_1^3; \\
    &\vdots \\
    vii) \ y_2^1 < y_1^2 < y_2^2 &< y_1^3 < y_1^1 < y_2^3; \\
    &\vdots \\
    xxx) \ y_2^3 < y_1^3 < y_2^1 &< y_1^2 < y_2^2 < y_1^1.
\end{split}
\end{equation}
By examining the relative positions of $y_1^2, y_1^3, y_2^2$, and $y_2^3$ (which constitute $w_1^2$) in each of the former 30 inequalities, we can ultimately assign an allowed permutation pattern $\Pi$ to $\pi_1^2$: \textit{i)} $\Pi = (0,2,1,3)$; \textit{ii)} $\Pi = (0,2,3,1)$; $\dots$; \textit{vii)} $\Pi = (0,2,1,3)$; $\dots$; \textit{xxx)} $\Pi = (3,1,0,2)$. As we have previously discussed, there are only $12$ unique permutations that can horizontally follow $\pi_1^1$, meaning that some permutations associated with the previous 30 inequalities appear more than once. These different frequencies of occurrence will end-up implying the existence of different edge weights in random ordinal networks. 

By repeating the same procedure for
$w_1^1 = 
\begin{pmatrix} y_1^1 & y_1^2 \\
y_2^1 & y_2^2
\end{pmatrix}$ 
and
$w_2^1 = 
\begin{pmatrix} y_2^1 & y_2^2 \\
y_3^1 & y_3^2
\end{pmatrix},$
we find another set of 30 inequalities and their corresponding permutations. Thus, from the analysis of amplitude relations between data values in $w_1^1$ and its neighbors $w_1^2$ and $w_2^1$, we find a total of 60 inequalities corresponding to 60 non-unique permutations. Because amplitude relations in all these inequalities involve random data, all these 60 inequalities are equiprobable~\cite{pessa2019characterizing}, and we can count the number of unique permutations (stemming from $\pi_1^1$) to define their relative frequencies. Finally, we normalize these transition probabilities from $\pi_1^1$ to all allowed permutations (edge weights Eq.~\ref{eq:edge_weights}) at the node level by dividing the frequency of occurrence of each unique permutation by the total number of possible inequalities involving the elements of $w_1^1$ and its neighbors $w_1^2$ and $w_2^1$ (60 in case $d_x = d_y =2$). In addition, we also normalize the transitions at the network level by dividing all edge weights by $1/(d_xd_y)!$. This last step is necessary so that the out-strength of permutation $\pi_1^1$ reflects the fact that all different permutations occur with equal probability in random data~\cite{ribeiro2012complexity,bandt2002permutation}.

We have automatized the former procedure to explicitly consider each possible ordinal pattern in $w_1^1$ [that is, $\pi_1^1 = (0,1,2,3)$, $(0,1,3,2)$, $(0,2,1,3)$, and so on] and thus estimate the edge weights for all allowed transitions in an ordinal network mapped from a large sample of random data. This approach allows us to completely specify a random ordinal network for arbitrary embedding dimensions ($d_x$ and $d_y$) and estimate all its relevant network metrics (including the global node entropy defined in Eq.~\ref{eq:global_node_entropy}).

\subsection{Ordinal networks of noisy-periodic ornaments}\label{sec:ornaments}

\begin{figure*}
\centering
\includegraphics[width=1\linewidth]{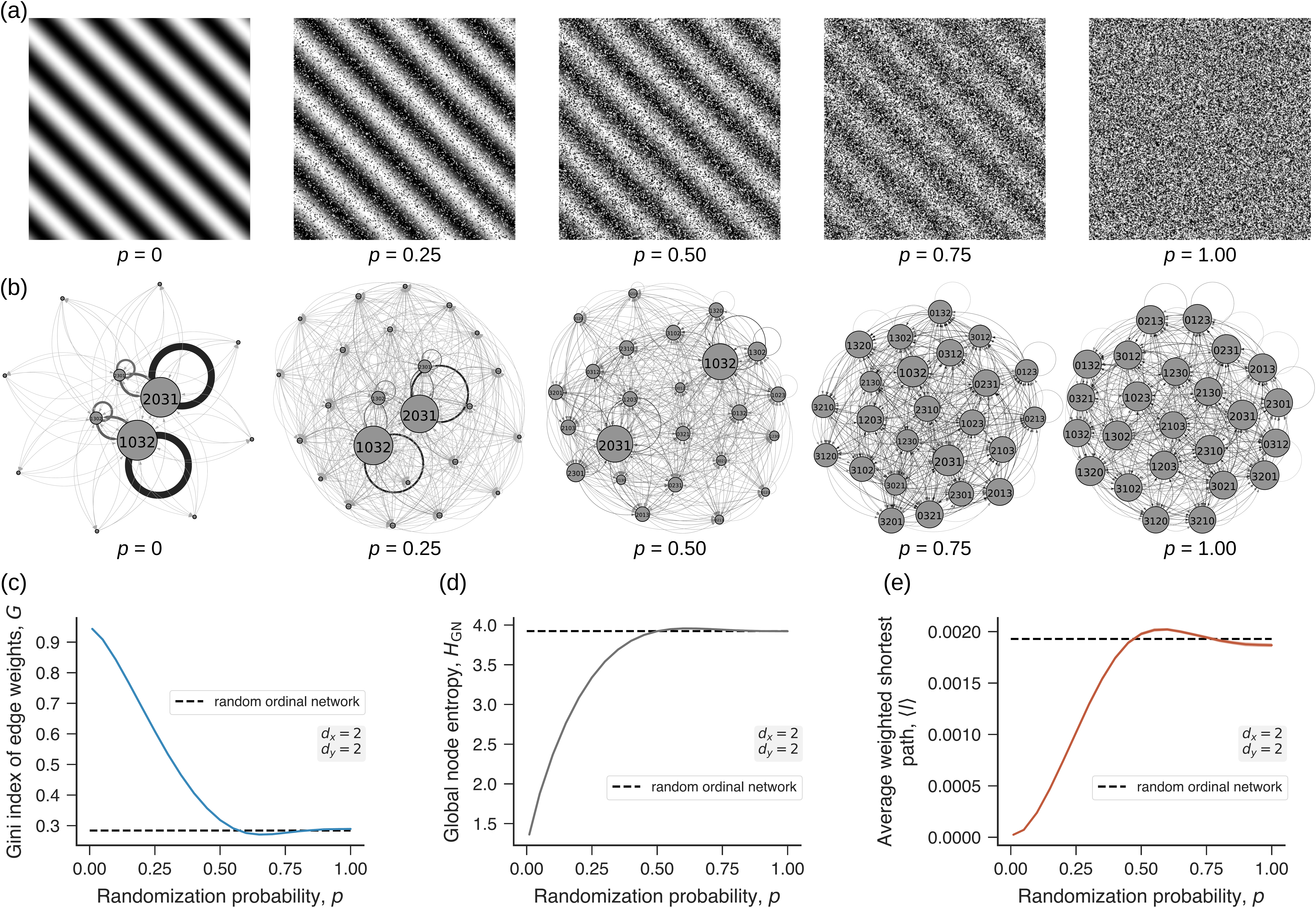}
\caption{Emergence of random ordinal networks in noisy-periodic ornaments. (a) Visualizations of geometric ornament images for different randomization probabilities $p$ (shown below images). (b) Ordinal networks with $d_x = d_y = 2$ mapped from ornament images displayed in the previous panel. The thicker and darker the edge, the higher is the edge weight (probability associated with permutation transition). Node sizes reflect total in-strength so that the bigger node the more frequently it is found in the symbolic sequence. Dependence of the (c) Gini index of edge weights $G$, (d) global node entropy $H_{\rm GN}$, and (e) average weighted shortest paths $\langle l \rangle$ on the randomization probability $p$. In the last three panels, the continuous lines show the average values of the network measures and the shaded regions indicate one standard deviation band estimated from an ensemble of $100$ ornament samples of size $250\times250$ for each value of $p$. The black dashed lines indicate the exact values of the network measures for random ordinal networks.}
\label{fig:ornament}
\end{figure*}

An interesting application to observe the emergence of random ordinal networks is the randomization process of a geometric ornament~\cite{zunino2016discriminating}. Figure~\ref{fig:ornament}{(a)} illustrates this procedure, where the probability of randomly shuffling pixels values $p$ controls the transition from a periodic image ($p=0$) to a random one ($p=1$). We map these sample images (of size $250\times250$) into their corresponding ordinal networks for $d_x = d_y = 2$, as shown in Fig.~\ref{fig:ornament}{(b)}. The visual aspect of these ordinal networks alludes to the visual features of the noisy-periodic ornaments. For small values of $p$, we observe that the ordinal networks are marked by intense connections between nodes $(1,0,3,2)$ and $(1,3,0,2)$ as well as between $(2,0,3,1)$ and $(2,3,0,1)$, and by auto-loops in these four nodes, which in turn reflect the diagonal stripes in the images. As the values of $p$ increases, the strength of these connections fade out and give rise to a more uniform distribution of edge weights. 

To systematically investigate these noisy-periodic ornaments, we generate an ensemble containing 100 ornament samples of size $250\times250$ for each randomization probability $p\in\{0.01,0.05,0.10,\dots,1.00\}$ and transform them into ordinal networks using embedding parameters $d_x = d_y = 2$. From these networks, we estimate the average value of the Gini index of edge weights $G$, the global node entropy $H_{\rm GN}$ (see Eq.~\ref{eq:global_node_entropy}), and the average weighted shortest path $\langle l \rangle$, as shown in Figs.~\ref{fig:ornament}{(c)-(e)}. The Gini index is defined as~\cite{cowell2011measuring}
\begin{equation}
G = \frac{\sum_i^{N_e}(2i-N_e-1) x_i}{N_e\sum_{i}^{N_e}{x_i}} \,,
\end{equation}
where $x_i$ represents the edge weights in ascending order, $i$ is the edge weight rank, and $N_e$ is number of edges in the ordinal network. In weighted networks, the shortest path between two nodes is defined as the path minimizing the sum of the edge weights between them~\cite{newman2010networks}, and the average weighted shortest path is defined as the average value of the weighted shortest path among all pairs of nodes. In addition, we generate the exact form of the random ordinal networks for $d_x=d_y=2$ by following the prescriptions given in Section~\ref{sec:random} to estimate the values of these three network properties [dashed lines in Figs.~\ref{fig:ornament}{(c)-(e)} presented in Sec.~\ref{sec:random}]. 

As expected, we observe that the three measures approach the exact values for random networks with the increase of $p$. It is interesting to notice that the global node entropy $H_{\rm GN}$ surpasses the exact random value around $p\approx0.5$ before converging to it. This happens because random ordinal networks are not the most entropic ordinal networks~\cite{pessa2019characterizing}, a characteristic that is explained by the fact that edges weights are not all equal in random ordinal networks. We also observe that the average weighted shortest path $\langle l \rangle$ appears to converge to a value slightly lower than the expected from the exact form of the random ordinal network. This apparent discrepancy occurs because the relatively small size of the images prevents a more accurate estimation of all permutation transitions; however, this difference between the values of $\langle l \rangle$ for $p=1$ vanishes as we increase the image size.

\subsection{Ordinal networks of fractional Brownian landscapes}\label{sec:fbm}

\begin{figure*}[!ht]
\centering
\includegraphics[width=1\linewidth]{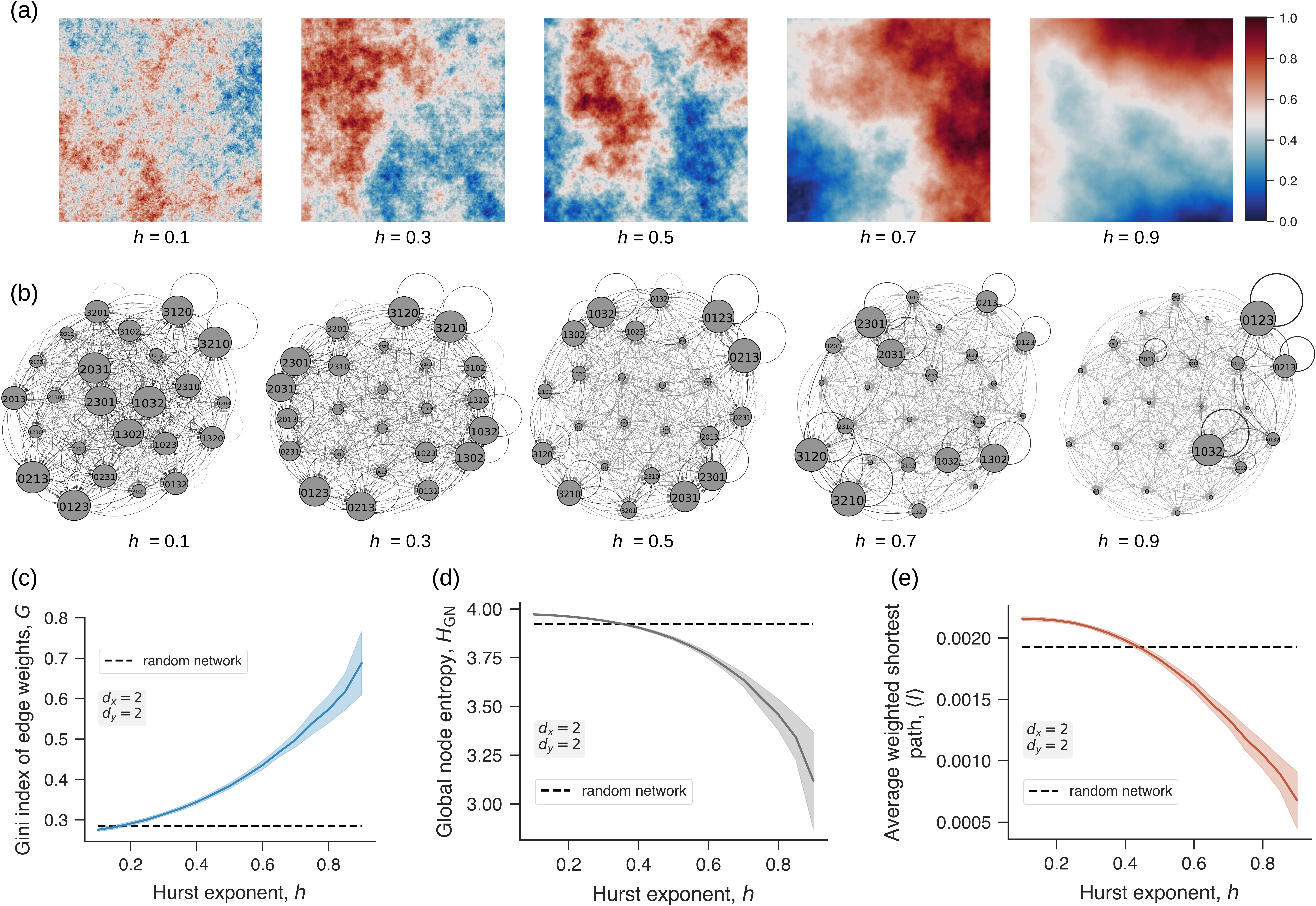}
\caption{Ordinal networks of fractional Brownian landscapes. (a) Examples of fractional Brownian surfaces for a few values of the Hurst exponent (shown below each image). We have normalized all surfaces so that blue shades indicate low height regions and red shades the opposite. (b) Ordinal networks mapped from the fractal surfaces shown in the previous panel. Dependence of the (c) Gini index of edge weights $G$, (d) global node entropy $H_{\rm GN}$, and (e) average weighted shortest paths $\langle l \rangle$ on the Hurst exponent $h$ of two-dimensional fractional Brownian motion. In these last three panels, the continuous lines represent average values (from an ensemble of 100 landscape samples for each $h$), and shaded regions stand for one standard deviation band. The black dashed horizontal lines represent the values of these network metrics estimated from random ordinal networks.}
\label{fig:fbm}
\end{figure*}

In another application, we investigate ordinal networks mapped from two-dimensional fractional Brownian motion~\cite{mandelbrot1982fractal}. This class of stochastic processes models natural landscapes and is characterized by the Hurst exponent $h\in(0,1)$ that controls the surface roughness. Surfaces generated with small values of $h$ ($h\to0$) are rough while large values of $h$ ($h\to1$) produce smooth landscapes. Cross sections of fractional Brownian landscapes with $h=1/2$ represent usual random walks or Brownian motion. Figure~\ref{fig:fbm}{(a)} shows examples of fractional Brownian landscapes generated by the turning bands method~\cite{yin1996turning} for different values of the Hurst exponent $h$.

We generate an ensemble containing 100 fractional Brownian landscapes of size $256 \times 256$ for each value of $h\in\{0.10,0.15,0.20,\dots,0.90\}$ (with the turning bands method), and map each sample into an ordinal network with embedding dimensions $d_x = d_y = 2$. Figure~\ref{fig:fbm}(b) presents visualizations of the ordinal networks mapped from the sample images of Fig.~\ref{fig:fbm}(a). We observe that changes in surface roughness affect the connectivity patterns of the resulting networks. Rougher surfaces produce ordinal networks with a more even distribution of edge weights which visually resemble random ordinal networks [last panel in Fig.~\ref{fig:ornament}{(b)}]. As fractional Brownian surfaces become smoother, we observe a concentration of weight in a few edges among particular nodes, while the intensity of most links decreases. This concentration of weight reflects the predominant occurrence of only a few permutations in the symbolic arrays related to smoother images.

\begin{figure*}[!ht]
\centering
\includegraphics[width=1\linewidth]{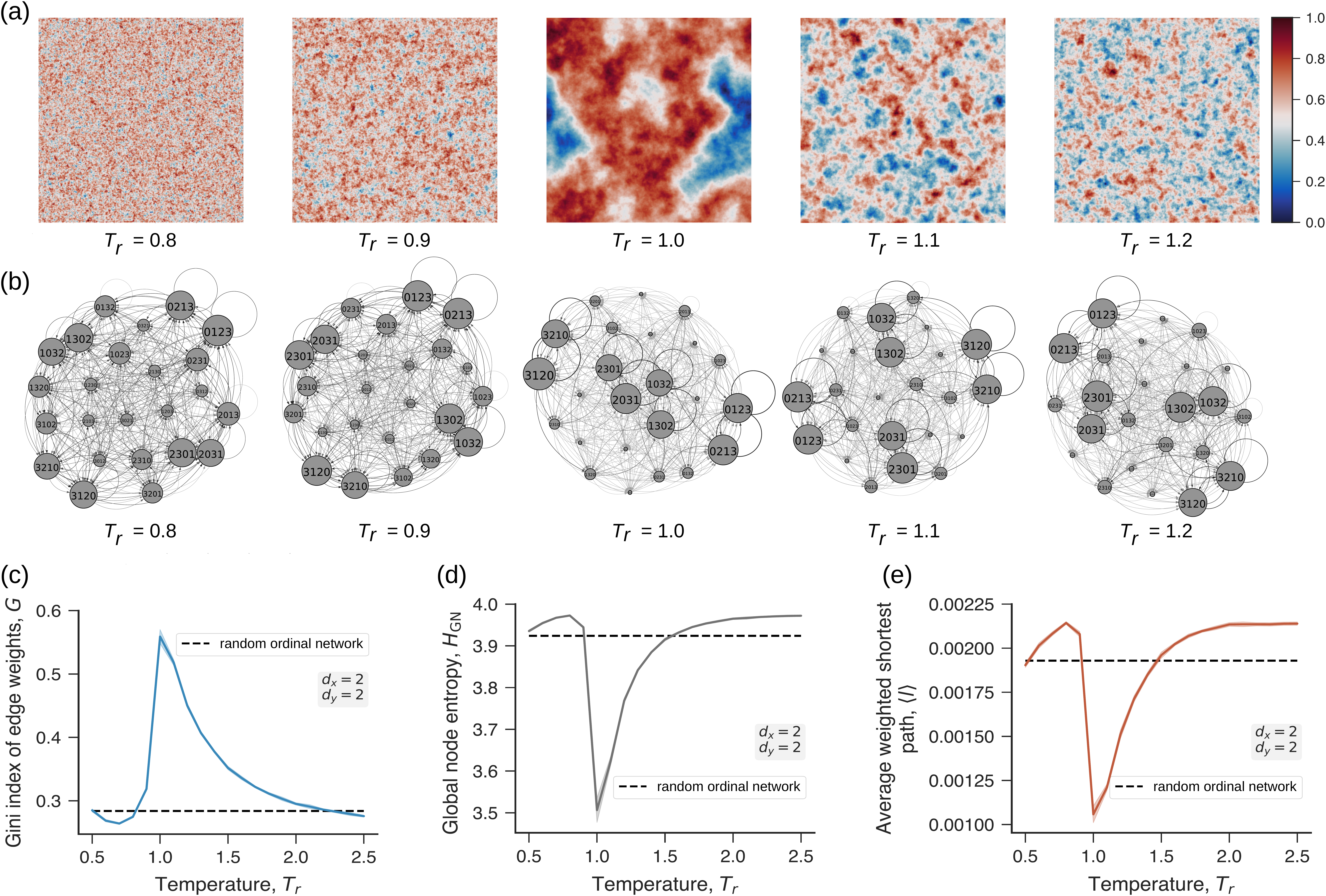}
\caption{Ordinal networks of Ising surfaces. (a) Illustration of Ising surfaces for different values of reduced temperatures $T_r$ (indicated below each image). We note that more complex patterns emerge around criticality ($T_r=1$). (b) Examples of ordinal networks mapped from the images in the previous panel. Dependence of the (c) Gini index of edge weights $G$, (d) global node entropy $H_{\rm GN}$, and (e) average weighted shortest path $\langle l \rangle$ on the reduced temperature $T_r$. The horizontal dashed lines in the previous three panels indicate values of the corresponding metric estimated from random ordinal networks.}
\label{fig:ising}
\end{figure*}

Beyond the previous qualitative observations, we calculate the average values of the Gini index of edge weights $G$, the global node entropy $H_{\rm GN}$, and the average weighted shortest path $\langle l \rangle$ as a function of the Hurst exponent $h$ using our ensemble of fractional Brownian landscapes. Figures~\ref{fig:fbm}(c)-(e) show these three network measures in comparison with their values estimated from random ordinal networks. These results are in line with our qualitative observations. Specifically, the Gini index increases with the Hurst exponent, reflecting the concentration of weight in fewer edges. On the other hand, the global node entropy $H_{\rm GN}$ and the average weighted shortest path $\langle l \rangle$ decrease as the Hurst exponent increases. The monotonic relationships of these three metrics as functions of the Hurst exponent indicate that their values are good predictors of landscape roughness. We have also verified that the behavior of these three network measures obtained for embedding dimensions $d_x = d_y = 2$ are very similar to those obtained for $d_x=3$ and $d_y=2$ (or $d_x=2$ and $d_y=3$). Nevertheless, we need larger fractional Brownian surfaces to proper estimate all transition probabilities.

\subsection{Ordinal networks of Ising surfaces}\label{sec:ising}

We have also applied ordinal networks to Ising surfaces~\cite{brito2007Dynamics,brito2010two} to verify whether network measures are capable of identifying phase transitions. These surfaces are obtained by accumulating the spin variables $\sigma(t)$ of the canonical Ising model in a Monte Carlo simulation~\cite{landau2015guide}. To describe this model, we consider a square lattice whose sites are occupied by spin-$1/2$ particles [$\sigma\in \{-1,1\}$] with Hamiltonian given by
\begin{equation}
    \mathcal{H} = -\sum_{\langle i,j \rangle}\sigma_i \sigma_j\,,
\end{equation}
where the summation is over all pairs of first neighbors. The height $S_i$ of the corresponding Ising surface at site $i$ is then defined as
\begin{equation}
    S_i = \sum_{t} \sigma_i(t)\,,
\end{equation}
where $\sigma_i(t)$ is the spin value in step $t$ of the Monte Carlo simulation.

Figure~\ref{fig:ising}{(a)} shows examples of Ising surfaces of size $250 \times 250$ obtained for different reduced temperatures $T_r = T/T_c$, where $T_c = 2/\ln(1+\sqrt{2})$ is the critical temperature at which the Ising model undergoes a phase transition. Surfaces generated at reduced temperatures distant from the critical value ($T_r=1$) do not exhibit long-range structures and are similar to two-dimensional white noise. However, we start to observe more complex patterns as the reduced temperature gets closer, and especially when it is equal, to the critical value.

We generate an ensemble containing $10$ Ising surfaces of size $250 \times 250$ for each value of $T_r\in\{0.5,0.6,\dots,2.5\}$. Each surface is obtained after accumulating the spin variables during 30,000 Monte Carlo steps to achieve equilibrium~\cite{brazhe2018shearlet,ribeiro2012complexity}. Next, we map all surfaces into ordinal networks with embedding dimensions $d_x = d_y = 2$. Figure~\ref{fig:ising}{(b)} shows examples of networks mapped from the images in Fig.~\ref{fig:ising}{(a)}. A visual inspection of these ordinal networks already suggests that their properties change with the reduced temperature.

\begin{figure*}[!ht]
\centering
\includegraphics[width=0.7\linewidth]{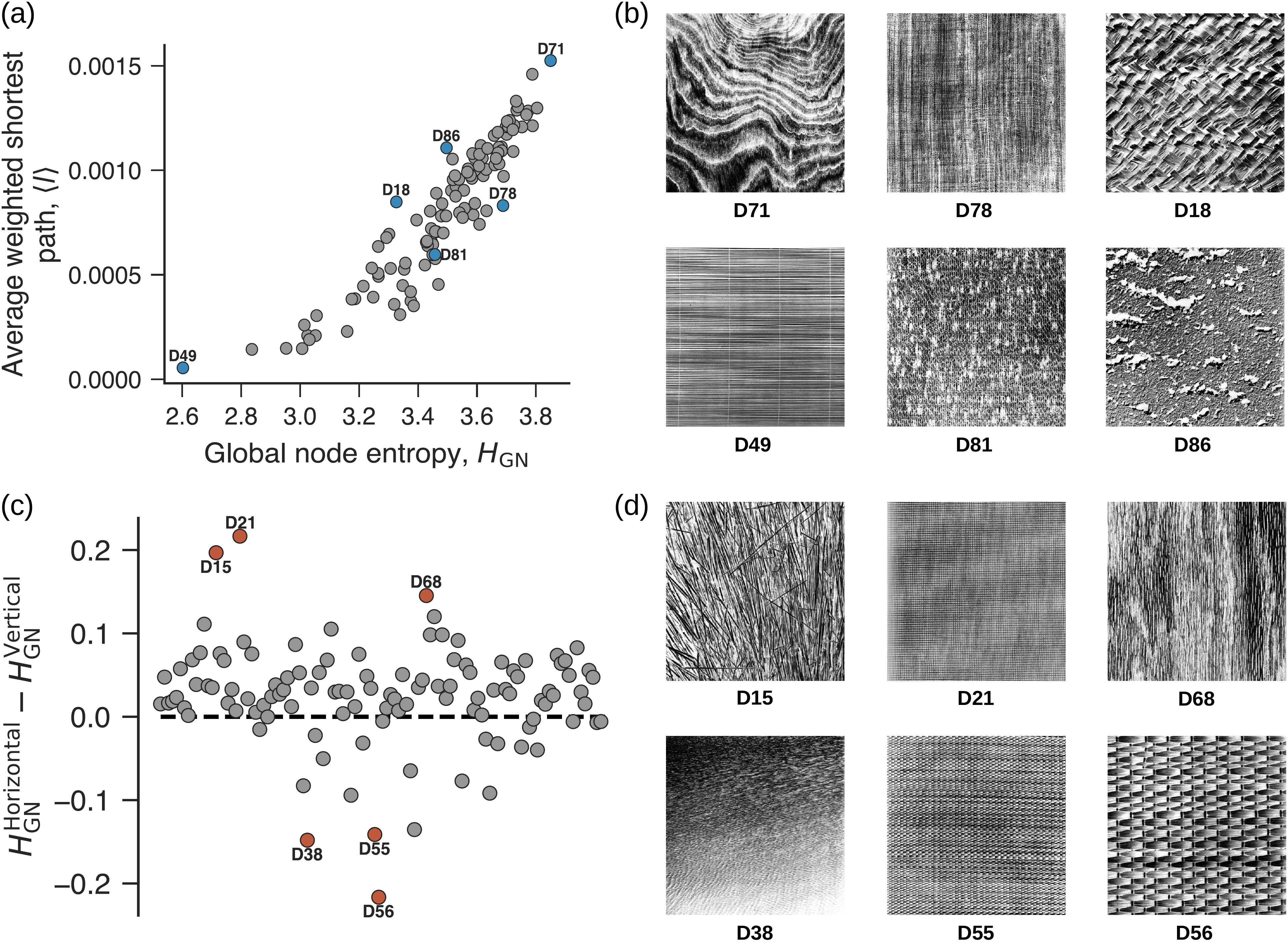}
\caption{Ordinal networks mapped from Brodatz textures. (a) Locations of all Brodatz textures at the plane of global node entropy $H_{\rm GN}$ versus average weighted shortest path $\langle l \rangle$. (b) Six different images corresponding to highlighted textures in the previous panel (blue markers). (c) Difference between the global network entropy estimated from horizontal and vertical ordinal networks ($H_{\rm GN}^{\rm Horizontal} - H_{\rm GN}^{\rm Vertical}$). (d) Six pictures corresponding to highlighted textures in the previous panel (red markers).}
\label{fig:brodatz}
\end{figure*}

Similarly to the previous applications, we calculate the Gini index of edge weights $G$, global node entropy $H_{\rm GN}$, and average weighted shortest path $\langle l \rangle$ as a function of the reduced temperature $T_r$. Results presented in Fig.~\ref{fig:ising}{(c)-(e)} show that all three measures present marked variations before and after the phase transition at $T_r = 1$, assuming extreme values at the critical temperature. The values at the critical temperature are also the furthest from those estimated from random ordinal networks [dashed lines in Fig.~\ref{fig:ising}{(c)-(e)}]. In addition, we note that variations of these metrics are steeper before than after criticality, indicating that spatial correlations are slowly broken with the rise of temperature. It is also worth observing that these networks become slightly more entropic than random ordinal networks at high temperatures.

\subsection{Ordinal networks of texture images}\label{sec:brodatz}

We also use our generalized algorithm to map real-world images to ordinal networks. To do so, we consider 112 pictures of natural textures comprising the Normalized Brodatz Texture database~\cite{safia2013new}, a set of 8-bit (256 gray levels) images of size $640 \times 640$ covering a large variety of natural textures that is often used as validation set for texture analysis~\cite{safia2013new}. This data set is an improvement over the original Brodatz album~\cite{brodatz1966textures} where background information is removed and the gray levels are uniformly distributed over the whole range. Thus, it is not possible to discriminate normalized Brodatz textures with simple statistical summaries based on gray-level distribution.

We map each image into an ordinal network with embedding dimensions $d_x = d_y = 2$. We then estimate the Gini index of edge weights $G$, global node entropy $H_{\rm GN}$, and average weighted shortest path $\langle l \rangle$ of each network. These network measures are cross-correlated with Pearson correlation coefficients ranging from $0.92$ for $\langle l\rangle$ versus $H_{\rm GN}$ to $-0.97$ for $G$ versus $H_{\rm GN}$. Figure~\ref{fig:brodatz}{(a)} shows a scatter plot of $\langle l\rangle$ versus $H_{\rm GN}$ (the less significant association) for all textures, where the dispersion pattern suggests that both measures are non-linearly related. Figure~\ref{fig:brodatz}{(b)} depicts six different images that are also highlighted in Fig.~\ref{fig:brodatz}{(a)}. We observe that the two textures with extreme values of $\langle l\rangle$ and $H_{\rm GN}$ (D49 and D71) are quite different: while regular horizontal stripes mark texture D49, texture D71 shows much more complex structures. 

We also note the existence of textures with similar values of $H_{\rm GN}$ and distinct values of $\langle l\rangle$ as well as images with similar values of $\langle l\rangle$ and different values of $H_{\rm GN}$. These results suggest that both measures may quantify different aspects of images. For instance, textures D86 and D81 [highlighted in Fig.~\ref{fig:brodatz}{(a)}] have almost the same values of global node entropy; however, the value of $\langle l\rangle$ is considerably larger for D86 than D81. By inspecting these two textures in Fig.~\ref{fig:brodatz}{(b)}, we note that D86 appears to be rougher than D81. Similarly, textures D18 and D78 have comparable values of $\langle l\rangle$ but quite different values of $H_{\rm GN}$. The visual inspection of these textures suggests that D18 is more regular and structured than D78. While it is challenging to generalize these interpretations to other images, we believe the values of $H_{\rm GN}$ quantify patterns at a more local level while $\langle l\rangle$ and $G$ are a more global measures. This idea somehow agrees with the definition of these measures in the sense that $H_{\rm GN}$ is based on relations involving first-neighbors, while $\langle l\rangle$ and $G$ involve the entire ordinal network.

We further investigate the possibility of exploring visual symmetries in the Brodatz data set. To do so, we have made a small modification in our original algorithm to create two ordinal networks from a single image. One of these networks considers only horizontal transitions among permutations (horizontal ordinal network), and the other uses solely vertical transitions among permutations (vertical ordinal network). Thus, we map each Brodatz texture into one horizontal and one vertical ordinal network with embedding dimensions $d_x = d_y = 2$. Then, we estimate the global node entropy from the horizontal network ($H_{\rm GN}^{\rm Horizontal}$) and the vertical network ($H_{\rm GN}^{\rm Vertical}$) by using Eq.~(\ref{eq:global_node_entropy}).

Figure~\ref{fig:brodatz}{(c)} shows the difference between these quantities ($H_{\rm GN}^{\rm Horizontal}-H_{\rm GN}^{\rm Vertical}$) for each texture. We observe a few textures with extreme values for this difference and highlight six of them, which are also depicted in Fig.~\ref{fig:brodatz}{(d)}. Most of these images are characterized by stripes or line segments predominantly oriented in vertical or horizontal directions, indicating that vertical and horizontal ordinal networks are capable of detecting this simple symmetry feature. These results indicate potential applications of our approach in classification tasks as a way of extracting texture features via network metrics~\cite{iacovacci2019visibility}, in a similar manner to which permutation based metrics have already been successfully used as predictors in image classification and regression problems~\cite{sigaki2018history,sigaki2019estimating}.

It is worth noticing that the pixel values of normalized Brodatz textures are discrete and may lead to equal values within the data partitions $w$ used for defining the ordinal patterns (Eq.~\ref{eq:partition}). As previously mentioned, we deal with these draws by keeping the occurrence order of the elements in $w$. This approach is common for ordinal methods in general, but there are other possibilities. For instance, the seminal work of \citeauthor{bandt2002permutation}~\cite{bandt2002permutation} proposes to add a small noise signal to data to remove possible equalities, while \citeauthor{bian2012modified}~\cite{bian2012modified} explicitly consider the equalities by repeating the permutation symbols. \citeauthor{zunino2017permutation}~\cite{zunino2017permutation} have found that these equalities can bias the value of permutation entropy of random time series, especially for large embedding dimensions, quite discrete signals, and small time series. This issue can also affect ordinal networks mapped from images with very low color depth. In these cases, one may need to consider different approaches for dealing with signal equalities. However, we have verified that our results for the Brodatz textures are not significantly affected by these equalities and are robust against adding a small noise signal to these textures.

\subsection{Comparison with traditional methods and robustness against noise}\label{sec:comparing}

As a final application, we compare image quantifiers derived from ordinal networks with those obtained from traditional texture descriptors. Specifically, we use the gray-level co-occurrence matrices (GLCM)~\cite{haralick1973textural,haralick1979statistical}, which is one of the most well-known and extensively used approaches for texture analysis~\cite{nanni2013different,cavalin2017review}. The elements of a GLCM represent the number (or probability) of gray-levels co-occurrence at a given distance ($D$) and angle ($\theta$) among the image pixels. In this approach, we obtain image features by calculating different statistical measures from the GLCM. Here, we use five standard properties of a GLCM (contrast, dissimilarity, homogeneity, energy, and correlation -- see Appendix~\ref{sec:GLCM} for definitions) for different values of $D$ and $\theta$ as implemented in the Python module scikit-image~\cite{van2014scikit}. We also consider three additional image quantifiers obtained from ordinal networks (strength variance, strength kurtosis, and weighted transitivity).

\begin{figure*}[!ht]
\centering
\includegraphics[width=1\linewidth]{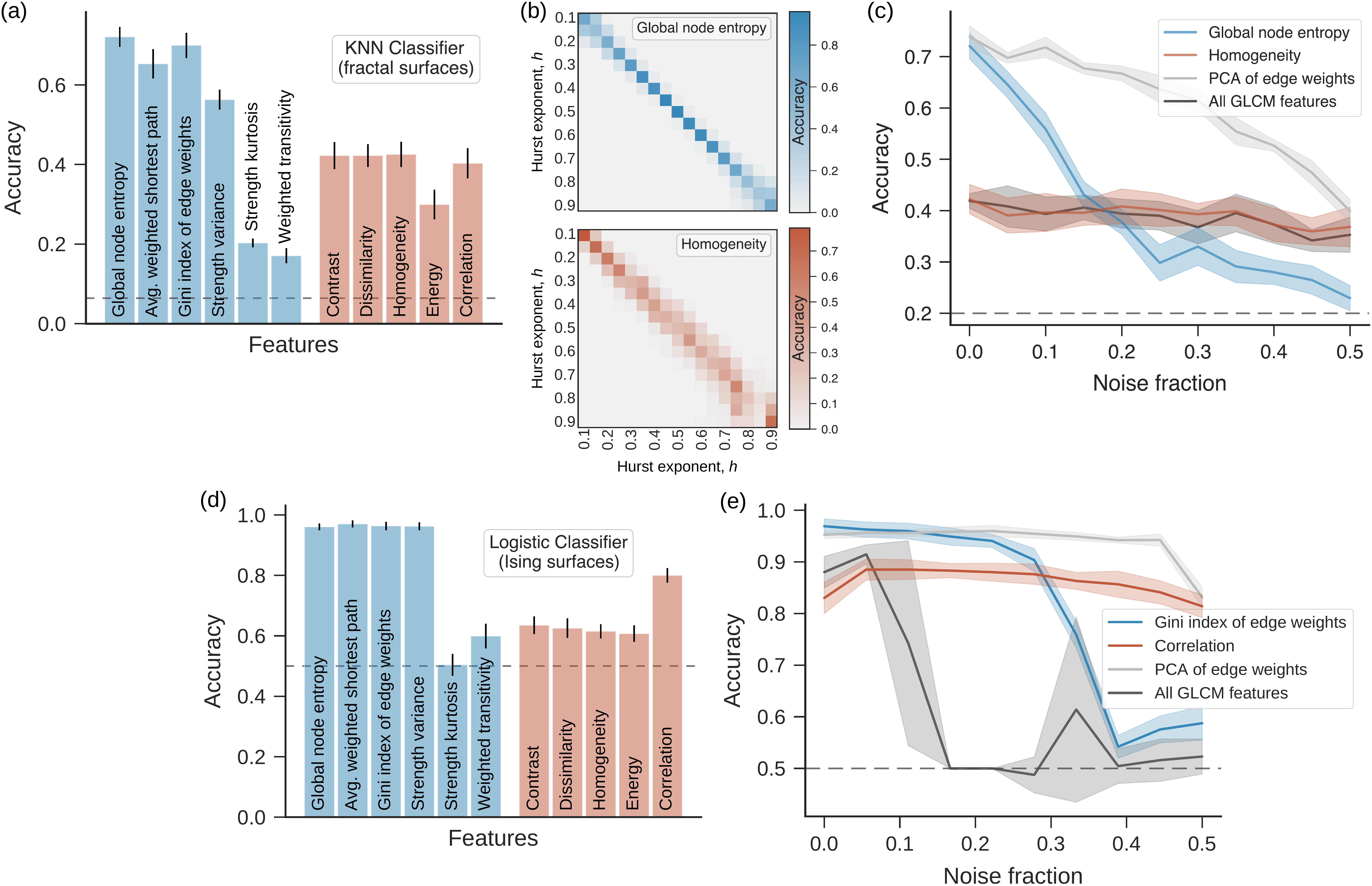}
\caption{Comparison with other approaches and robustness against noise. (a) Bars show the accuracy (fraction of correct classifications) for the classification task of predicting the Hurst exponent based on image features obtained from ordinal networks (blue bars) and the gray-level co-occurrence matrices (GLCM -- red bars). (b) True and predicted Hurst exponent. This confusion matrix details the performance of global node entropy in comparison with the best feature obtained from the GLCM approach (homogeneity). (c) Classification scores when using the Gini index of edge weights, GLCM correlation, five PCA components of the edges weights, and all five GLCM features for predicting the Hurst exponent as a function of the noise fraction added to the fractional Brownian landscapes. (d) Bars show the accuracy for the tasks of predicting whether Ising surfaces are above ($T_r=1.01$) or below ($T_r=0.99$) the critical temperature using image qualifiers obtained from ordinal networks (blue bars) and GLCM (red bars). (e) Accuracy of the classification tasks related to the Ising surfaces as a function of the noise fraction added to these images when using Gini index of edge weights, GLCM correlation, the first PCA component of the edges weights, and all five GLCM features. In all panels, the dashed lines indicate the baseline accuracy while error bars and shaded regions stand for one standard deviation.}
\label{fig:comparing}
\end{figure*}

We propose two classification tasks based on the numerical experiments previously described to compare the performance of image features derived from ordinal networks and GLCM. In the first task, we use these image features for predicting the Hurst exponent $h$ of fractional Brownian landscapes. To do so, we generate a data set comprising 100 fractal surfaces for each $h\in\{0.10, 0.15, \dots, 0.90\}$, and estimate the image quantifiers for both the GLCM and the ordinal network approaches. We train $k$-nearest neighbors algorithms~\cite{hastie2013elements} (as implemented in the Python package scikit-learn~\cite{pedregosa2011scikit}) for classifying the Hurst exponents in a 10-fold-cross-validation splitting strategy while optimizing the number of neighbors $k$ (the only parameter of the algorithm). In addition, we also select the optimal combination of $D\in\{1,2,3,4,5\}$ and $\theta\in\{0,45,90,135\}$ (often considered among the best setup~\cite{cavalin2017review}) for features extracted from GLCM. Results in Fig.~\ref{fig:comparing}(a) show that image features derived from ordinal networks display the top-4 best accuracies, with global node entropy being the best among all used features. The accuracy obtained for the best ordinal network features ($\sim\!70\%$) is considerably larger than the best GLCM features ($\sim\!40\%$). Figure~\ref{fig:comparing}(b) shows the confusion matrices of the classification tasks when using global node entropy (from ordinal network) and homogeneity (from GLCM) as features. The diagonal band in these matrices indicates that when algorithms incorrectly classifies the Hurst, they tend to predict a value that is close to actual Hurst of the fractal surface. However, the diagonal band observed for global node entropy is much narrower than the one observed for homogeneity, particularly for intermediate values of $h$.

In addition to being accurate, image quantifiers extracted from ordinal networks also need to be robust against noise, as images in practical applications usually have some degree of noise. To investigate this question, we consider again the classification experiments of fractal surfaces, but this time we add a fraction of ``salt and pepper'' noise (that is, a fraction of surface sites randomly selected have their values replaced by the maximum and minimum values of the surface) to the generated surfaces and estimate the classifier accuracy as a function of the noise fraction. Figure~\ref{fig:comparing}(c) shows this analysis when using global node entropy and homogeneity as classifier features. We observe that global node entropy outperforms the baseline accuracy ($20\%$) even with 50\% noise. However, the homogeneity feature is more stable against noise and starts to outperform global node entropy with $\approx20\%$ of noise. We also consider all five GLCM features together in this analysis, and the results of Fig.~\ref{fig:comparing}(c) indicate that the inclusion of more GLCM features improves neither the accuracy nor the noise robustness of the classifier. Conversely, if we use the five principal components obtained by applying principal component analysis (PCA)~\cite{hastie2013elements} to all edge weights of ordinal networks (only to the training set), we observe a slight improvement in the accuracy without noise and a considerable improvement in noise stability of the classifier. Indeed, these five PCA components significantly outperform the baseline and GLCM accuracies even with 50\% noise.

In the second classification task, we use image features from ordinal networks and GLCM to predict whether Ising surfaces (see Section~\ref{sec:ising}) are generated from temperatures below ($T_r<1$) or above ($T_r>1$) the critical temperature ($T_r=1$). To do so, we generate a data set comprising 1,000 Ising surfaces at the temperature $T_r=0.99$ and another 1,000 surfaces at $T_r=1.01$. Similarly to the first classification task, we train a logistic classifier to this binary classification problem and estimate the accuracy of the method when using the same five features extracted from ordinal networks and GLCM. Figure~\ref{fig:comparing}(d) shows that the top-4 best accuracies (about 95\%) are obtained with image qualifier extracted from ordinal networks, while GLCM correlation is the fifth-best feature with significantly lower accuracy (about 80\%). 

We have also studied how the accuracies of these binary-classification tasks are affected by noise addition. Figure~\ref{fig:comparing}(e) shows the accuracy when using Gini index of edge weights and GLCM correlation as a function of the fraction of ``salt and pepper'' noise added to the Ising surfaces. The results are similar to what we have found for the fractal surfaces. The Gini index of edge weights outperforms the baseline accuracy up to 50\% noise, but the GLCM correlation is more stable against noise addition and outperforms the ordinal network feature for more than 30\% of noise. When including all GLCM features as predictors, we observe a slight increase in accuracy (for small noise fractions) compared with the case where only GLCM correlation is used; however, this makes the approach very unstable against noise [dark gray curve in Fig.~\ref{fig:comparing}(e)]. On the other hand, the first PCA component obtained from the edge weights of ordinal networks significantly improves the noise stability of the logistic classifier.

\section{Conclusion}\label{sec:conclusion}

We have proposed a generalization of the ordinal network algorithm for mapping images (two-dimensional data) into networks. After describing the method, we have studied basic connectivity patterns of these networks which in turn allowed us to find the exact form of ordinal networks mapped from random data. We have observed the emergence of these random ordinal networks in a controlled setting by randomizing a geometric periodic ornament. We have also investigated changes in surface roughness of two-dimensional fractional Brownian motion and found monotonic relations between network measures and the Hurst parameter of these fractal surfaces. This result extends similar findings previously obtained from time series data~\cite{pessa2019characterizing}. In the context of physical models, we have studied phase transition in an Ising-like model where variations in network metrics have been proven useful for accurately identifying the critical temperature. We have also mapped natural texture images into ordinal networks and considered the possibility of using networks to potentially extract image features. Finally, we have compared the performance of image quantifiers extracted from ordinal networks with traditional texture descriptors obtained from the GLCM approach in more practical situations involving the classification of images. Our results demonstrate that ordinal networks can display a higher discrimination power than usual approaches while also being robust against noise addition to images. 

Ordinal networks can be thought of as an extension of the Bandt and Pompe~\cite{bandt2002permutation} symbolization approach, where not only the occurrence of ordinal patterns is investigated but also the relative frequency of the transitions among these patterns. Thus, these networks simultaneously encode statistics related to the ordinal probability distribution (the in-strengths) and information related to the transition probabilities. Furthermore, network measures estimated from ordinal networks combine these two information sources into single metrics that may present higher discrimination power when compared with summary statistics derived solely from the ordinal probability distribution (such as the permutation entropy and the statistical complexity~\cite{bandt2002permutation,rosso2007distinguishing,ribeiro2012complexity}).

Our work thus contributes to recent developments in network science focused on mapping two-dimensional data into networks and characterizing them through network representations~\cite{xiao2014row,lacasa2017visibility}. As these network approaches are quite novel when compared to other attempts stemming from complexity science~\cite{ribeiro2012complexity, zunino2016discriminating,brazhe2018shearlet,cai2006new,andrienko2000complexity,feldman2003structural}, we believe there are several opportunities for applying these new tools to different contexts involving image analysis as well as in practical applications involving image classification. We plan to release a Python module implementing our two-dimensional ordinal networks and other ordinal methods soon, but the current version of our code is available upon request.

\appendix
\section{Gray-level co-occurrence matrices}\label{sec:GLCM}
Gray-level co-occurrence matrices (GLCM)~\cite{haralick1973textural,haralick1979statistical} is a very popular technique for texture analysis. This approach consists of calculating matrices $m_{\theta,D}(i,j)$ representing the relative frequency of co-occurrence of pixel intensities $i$ and $j$ (with $i,j\leq N_g$, where $N_g$ is the number of gray levels of the image) at distance $D$ and angle $\theta$. Having the GLCM, one can use several statistical measures to describe the texture~\cite{haralick1979statistical}. We have used five standard measures as implemented in the Python module scikit-image~\cite{van2014scikit}. They are:
\begin{equation}
    \text{contrast}(\theta,D)=\sum_{i,j}^{N_g} (i-j)^2 m_{\theta,D}(i,j)\,,
    \vspace*{0.2cm}
\end{equation}
\begin{equation}
    \text{dissimilarity}(\theta,D)=\sum_{i,j}^{N_g}|i-j|m_{\theta,D}(i,j)\,,
\end{equation}
\begin{equation}
    \text{homogeneity}(\theta,D)=\sum_{i,j}^{N_g}\frac{m_{\theta,D}(i,j)}{1+(i-j)^2}\,,
\end{equation}
\begin{equation}
    \text{energy}(\theta,D)=\sqrt{\sum_{i,j}^{N_g} m_{\theta,D}(i,j)^2}\,,
\end{equation}
{\rm and}
\begin{equation}
    \text{correlation}(\theta,D) = \sum_{i,j}^{N_g} m_{\theta,D}(i,j) \frac{(i-\mu_x)(j-\mu_y)}{\sigma_x \sigma_y}\,,
\end{equation}
{\rm where} 
\begin{equation*}
\begin{split}
    \mu_x & = \sum_{i,j}^{N_g} i\; m_{\theta,D}(i,j)\,,\\
    \mu_y & = \sum_{i,j}^{N_g} j\; m_{\theta,D}(i,j)\,,\\
    \sigma_x^2 & = \sum_{i,j}^{N_g} (i-\mu_x)^2 m_{\theta,D}(i,j)\,,\\
    \sigma_y^2 & = \sum_{i,j}^{N_g} (j-\mu_y)^2 m_{\theta,D}(i,j)\,.
\end{split}
\end{equation*}

\bibliography{2d_ordinal_network}

\end{document}